# WHY SHOULD THE Q-METHOD BE INTEGRATED INTO THE DESIGN SCIENCE RESEARCH? A SYSTEMATIC MAPPING STUDY

*Research paper*

Irawan Nurhas, Institute for Positive Computing of Hochschule Ruhr West University of Applied Sciences, Bottrop, Germany, and University of Jyväskylä, Jyväskylä, Finland, irawan.nurhas@hs-ruhrwest.de

Stefan Geisler, Institute for Positive Computing of Hochschule Ruhr West University of Applied Sciences, Bottrop, Germany, stefan.geisler@hs-ruhrwest.de

Jan M. Pawlowski, University of Jyväskylä, Jyväskylä, Finland, and Hochschule Ruhr West University of Applied Sciences, Bottrop, Germany, jan.pawlowksi@hs-ruhrwest.de

## Abstract

*The Q-method has been utilized over time in various areas, including information systems. In this study, we used a systematic mapping to illustrate how the Q-method was applied within Information Systems (IS) community and proposing towards integration of Q-method into the Design Sciences Research (DSR) process as a tool for future research DSR-based IS studies. In this mapping study, we collected peer-reviewed journals from Basket-of-Eight journals and the digital library of the Association for Information Systems (AIS). Then we grouped the publications according to the process of DSR, and different variables for preparing Q-method from IS publications. We found that the potential of the Q-methodology can be used to support each main research stage of DSR processes and can serve as the useful tool to evaluate a system in the IS topic of system analysis and design.*

# 1      Introduction

In 1988, Kaplan and Duchon (1988) used the Q-method to validate user statements, in a case study related to user perceptions of the relation between work and the use of computer information systems. A premiere introduction of the Q-methodology to the Information Systems (IS) community was initiated in 2002 (Thomas & Watson, 2002). The authors introduced using the Q-method within IS studies by providing an example and several key points IS researchers must consider when implementing the Q-method. The Q-method still has not gained broad popularity within the IS community (Wingreen & Blanton, 2018). Nonetheless, there are IS-studies using the Q-method published in the Basket-of-Eight Journals (see Appendix 1), known as the best journals in the IS field (Lowry et al., 2013). IS researchers not only use the Q-method to identify a problem (Sutton, Khazanchi, Hampton, & Arnold, 2008) but also propose its use to evaluate a system adoption for better results based on user preference (Matzner, Hoffen, Heide, Plenter, & Chasin, 2015). Both problem identification and system evaluation are essential parts of the process in Design Science Research (DSR) (Peffers, Tuunanen, Rothenberger, & Chatterjee, 2007). They have already gained popularity, but researchers still seek "the missing link" they can use to support a social-technical view of design science (Carlsson, Henningsson, Hrastinski, & Keller, 2011). Therefore, this study aims to promote the use of the Q-method as a tool to support the DSR process.

The main goal of this paper is to elaborate on integrating the Q method into different phases of the Design Science Research process, to improve the result of subjectivity-driven Information Systems studies purposefully designed and evaluated by subjective human perception (Larsen, Sørebø, & Sørebø, 2009; Venkatesh & Davis, 2000), considering their significant influence on the design for both organizations and the information technology (Larsen et al., 2009). Therefore, integrating the subjectivity of the user experience with the system design process takes a vital role within IS studies (Schepers & Wetzels, 2007). User experience is one critical point in the design of systems. Currently, the focus is changing toward well-being (Pawlowski et al., 2015) and measuring subjectivity in personal experience (Calvo & Peters, 2014). User-system experiences differ; by correctly identifying the subjective requirements of a technological environment (Calvo & Peters, 2014), researchers can provide insight into sustainable use of a system for a particular user and context. This can help the organization choose the design of the system or evaluate the implemented system-design solution (Pawlowski et al., 2015).

In 1953, Stephenson developed the Q-method to measure subjectivity (Jung et al., 2009; Stephenson, 1953). The Q-method or Q-sort combines qualitative and quantitative approaches to measuring human subjectivity (Dziopa & Ahern, 2011; Jung et al., 2009; Newman & Ramlo, 2010). In other words, the Q-method tries to get the users' perceptions or opinions on the statements; then evaluate and order the statements by importance in a "forced-choice" distribution form known as Q-sort, in order to reach consensus (Dziopa & Ahern, 2011; Stephenson, 1953). The Q-method is already in extensive use in social-science research (Doody, Kearney, Barry, Moles, & O'Regan, 2009; Jung et al., 2009), to evaluate educational technology (Kurt & Yıldırım, 2018; Wharrad & Windle, 2010), theory building or theory testing (Yang, 2016), and proposed IS-design solutions (Doherty, 2012; Gottschalk, 2002; Mettler & Wulf, 2018). An active Q-method community has consistently organized conferences each year since 1983.[1]

In design science, three categories constitute the main sources of artifact theory or design: people, processes, and products (Cross, 1999). The first category reflects subjective opinions and perception of human experiences (Cross, 1999). Therefore, IS researchers should initiate integrating Q-method into design science as one tool for studying subjectivity (Thomas & Watson, 2002). A research essay (Oberländer, Röglinger, Rosemann, & Kees, 2018) in the *European Journal of Information Systems*

---

[1] https://qmethod.org/qconference/ last accessed on 29.01.2019

*(EJIS)* discusses the use of DSR process integrating Q-method (Hrastinski, Carlsson, Henningsson, & Keller, 2008). Unfortunately, the IS community lacks IS studies exploring the potential use of the Q-method for the DSR process. Therefore, this study pioneers the initial step toward the future integration of the Q-method in the DSR process by providing an overview of its current implementation in IS studies. This paper aims to answer the main research question: "Why should the Q-method be integrated into the DSR process?"

This study uses systematic mapping (Petersen, Feldt, Mujtaba, & Mattsson, 2008) because of the lack of suitable studies (Petersen et al., 2008), notably in the context of the Q-method in DSR processes. The systematic-mapping study comprises an option for systematic reviews (Petersen et al., 2008) and could be suitable for the condition where there is little or no empirical evidence to apply a systematic review directly (Engström & Runeson, 2011). A mapping study was performed with the aim of understanding research gaps and assembling evidence to motivate future research (Engström & Runeson, 2011). Therefore, despite systematic mapping requiring less effort than a literature review, it still provides a more granular overview (Barbosa, Santos, Alves, Werner, & Jansen, 2013) for study purposes. This study contributes to extending the use of the Q-method into the DSR process within the IS community, and provides an overview of optimizing the full potential of the Q-method, which has been implemented minimally, if at all, within IS studies so far.

This paper is structured as follows. First, the mapping process is provided, including the construction of the guideline questions, selection of publications, and data extraction. Next, the results are presented based on the selection process and the variables for classification in the guideline questions. Then, the results and paper contributions are discussed, leading finally to identifying limitations and further research directions.

## 2 The Systematic-Mapping Process

The systematic mapping for this study followed the guideline for administering a structured-mapping process (Petersen et al., 2008). The mapping process begins with the construction of a framework in the form of guideline questions. Then publications are chosen according to inclusion and exclusion criteria. The mapping process was conducted based on the extracted data and the paper classified by following the constructed guideline framework.

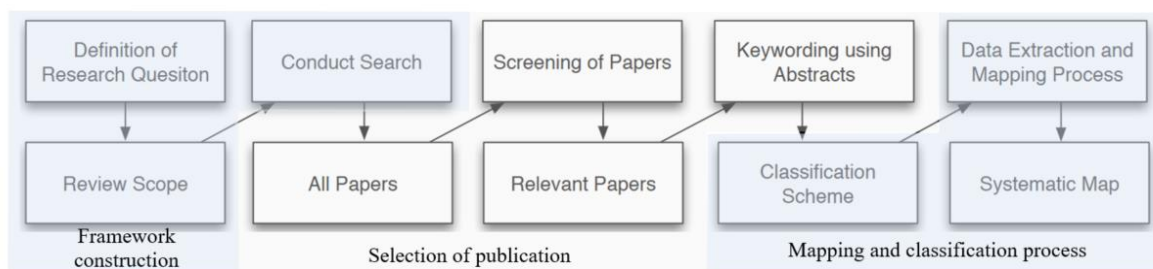

*Figure 1.    The systematic mapping process adopted from (Petersen et al., 2008)*

### 2.1    Construction of guideline framework for the mapping study

The explanation of the guideline questions as a framework for conducting the mapping process takes the major part of section 2. The guideline framework directs the grouping and data-extraction process that includes both searching strategies and the classification. The guideline questions are one of the sources for discussing and analyzing the contribution of this study. Therefore, the construction of the guideline questions also takes a role as background theory. In this study, the main topic is DSR and the Q-method. Then, guideline questions were constructed from both study topics by providing essential variables that allowed an overview and answering the main research question.

### 2.1.1 The Q-method for information systems studies

Extracting information related to the use of the Q-Method in IS publications provides an overview of gaps and opportunities for further Q-Method implementation. Publications on the use of the Q-method in IS studies were classified based on IS topics/thematic classification, to identify the possible fields of use. One difference between the systematic mapping and systematic review is the thematic classification of publications (Petersen et al., 2008). The thematic classification here followed core topics for the IS curriculum (Topi et al., 2010), including:

- IS foundation as an overview of the overall key topics in the field of IS.
- The data and information management cover study about knowledge management, database-management system, and business intelligence.
- The enterprise architecture.
- IS project management and Information Technology (IT) Infrastructure.
- The system analysis and design, covering an aspect of process design and analysis, evaluation of system, tools, method, and technique to identify requirements of IS and improve a system.
- The IS strategy, Management, and Acquisition.

The Q-method consists of *a concourse* or a list of statements, opinions, or the sum of all that people say or think about the investigated topic collected from various sources (Watts & Stenner, 2012; Yang, 2016). Statements are essential to preparation and can be used to represent the feelings and opinions of the user/object of study (Watts & Stenner, 2012). Preparing "the phenomena of mind" involves researchers who implement the Q-method, so the participants can easily understand (Watts & Stenner, 2012) the sorting process (Q-sort). The conventional way to collect the initial statements is interviewing or direct interaction with the system user (Wharrad & Windle, 2010), or by asking experts' opinions (Chang, 2012; Segars & Grover, 1998). Researchers can also collect statements from indirect interaction, such as through the literature-review process (Chang, 2012; Watts & Stenner, 2012). Alternatively, any digital sources (e.g., blogs, news websites, videos, images, social media status) (Davis & Michelle, 2011; Orchard, Fullwood, Morris, & Galbraith, 2015) or any type of big data could be used (Lynch, Adler, & Howard, 2014; Watts & Stenner, 2012). One of the strong points of the concourse in the Q-methods is the flexibility to collect the statements from various sources (Watts & Stenner, 2012). If a concourse is already prepared, it will be presented and sorted by using manual card sorting (Watts & Stenner, 2012), offline computerized software (Watts & Stenner, 2012) or an online system (Watts & Stenner, 2012). In the Q method, a study depends on human subjectivity and cannot be performed without it. Furthermore, providing a large number of respondents for the sorting of statements and the use of different tools is not the main focus (Stephenson, 1953; Watts & Stenner, 2012). Therefore, debate persists about the ideal number of participants in the sorting process of the concourse. The type of sources, the role of the Q-method in the studies, and the topic of publication influence the sorting process and analysis (Watts & Stenner, 2012), and can support understanding the current use of Q-method in the IS community. Therefore, understanding how those three variables were applied differently for each process of DSR leads to the Guideline Question (GQ) related to the Q-method implementation in IS studies: *GQ1) How do IS researchers use the Q-method?*

### 2.1.2 Design Science Research (DSR)

DSR is an iterative process for problem-based system design (Peffers et al., 2007). Its main aim is to design toward a perfect artifact by improving the status quo (Gregor & Hevner, 2013). DSR requires IS researchers to focus on a holistic view of artifact evaluation. The attention of IS scholars on the use of DSR for IS studies continues to grow and is now firmly established design process and method within the IS community (Prat, Comyn-Wattiau, & Akoka, 2014). Nonetheless, there are different genres of DSR (Iivari, 2015; Peffers, Tuunanen, & Niehaves, 2018) that require different understandings. IS researchers sometimes have difficulty identifying and presenting DSR study contributions (Peffers et al., 2018). This study's purpose is best seen as using the perspectival lens of DSR as a research process (Peffers et al., 2007) that can be started and deliver IS artifact at each stage of the DSR process (Peffers et al., 2007). From this perspective, Peffers et al. (2007) present six main processes for DSR, starting from problem identification and motivation, followed by the process to define the solution, the design and development of the system, the demonstration of the system in the form of a working prototype or a case study, system evaluation, and finally, the communication of the study result (Peffers et al., 2007) to get feedback and reflection for the iterative process. Hevner and Chatterjee (2010) simplify the whole process and conclude that IS scholars should develop IS artifacts by addressing the critical problems (problem-driven process), designing and demonstrating the artifacts (solution-driven process), and evaluating and predicting the potential benefits and risks of the proposed artifacts (evaluation-driven process). The three main processes of DSR serve for this study as the variables (the type of the process) within the process. The other classification of the variables is the contribution from the literature.

For DSR studies, Gregor and Hevner (2013) provide a knowledge-contribution framework by classifying the contribution into four groups. The DSR study can provide a new solution for a new problem (contribute as an invention) or for known problems (contribute to an improvement). In the mature solution, two more classifications were added; the exaptation of extending the known solution to new problems; and the study made no major contribution or only applied known solutions to known problems (Gregor & Hevner, 2013). Understanding the paper's contribution of applying the Q-method, as well as the process and activities within the DSR process (Baskerville, Pries-Heje, & Venable, 2009) can support analysing the integration of the Q-method into the DSR process. Accordingly, the next GQ is used for the mapping study: *GQ2) To which DSR processes can IS researchers apply the Q-methodology?*

## 2.2 Selection of publications

This section briefly explains the filtering of selected publications, including the inclusion and exclusion criteria. First is selecting Basket-of-Eight journals as primary sources for the journal collection. These journals are known as trusted and high-quality journals for IS scholars (Lowry et al., 2013). Also, automatic searches of the AIS Digital Library (AIS DL) specifically focusing on the publication database for IS studies were implemented. Using different keywords (e.g., "Q-method," "Q sort," "Q-methodology," "Q technique") on January 21, 2019, resulted in 47 publications from Basket-of-Eight journals and 211 publications from the ACM Digital Library. After implementing inclusion criteria (i.e., only journal publications, only research articles or literature-review articles) and removing duplicates, by applying to 70 publications the exclusion criteria (i.e., not written in English) and keyword location filters (i.e., keyword only found in the list of reference and not actually used in the study), the final 45 selected publications were identified.

## 2.3 Mapping and classification process

The classification and data-extraction process began with ordering the publications by year and journal name to find the distribution of the papers. As mentioned in the guideline framework, different variables shaped the classification (e.g., theme, source of the concourse, the comparable process of DSR, and the study contribution). Then, the classification related to the use of the Q-method of each publication was identified. Concerning the role of the Q-method in a publication, the Q-method can be used as the primary research method or as a supporting method to strengthen the main study method. The role of the Q-method was identified as the main method/review if it was the only method used in the study and is mentioned in the title, abstract, or keywords section. Moreover, use of the Q-method as the support method was identified as major support (if the author mentioned that the Q-method was used in the study with proper explanation and analysis) or minor support (without further detail, information, implementation, and analysis).

Furthermore, whether a paper provides a paragraph that explains the Q-method also aided in the classification based on the role of the Q-method. Three classifications of DSR process (i.e., problem identification, solution definition, and evaluation) and four classifications of knowledge contribution (i.e., no significant knowledge contribution, exaptation, improvement, and invention) were identified and analyzed for each publication. Classification in terms of the source of the concourse (expert opinion; user interviews; literature review; observation; not scientifically based digital media including tweets, post, images, or videos) was extracted from each paper and converted to the number of types of sources used, with or without literature. The result is presented in the following section.

## 3     Result

The publication date and the journal channel of each article create an overview of the paper distribution (see Appendix 1 and Appendix 2). Although the time interval for the year of publication was not limited, no publications used the Q-method before 1988 (e.g., *MISQ* has existed since 1977). Approximately 35 years elapse from the first introduction of the Q-method in 1953 (Stephenson, 1953) to first use of the method for IS study in 1988 (Kaplan & Duchon, 1988), and still only a limited number of IS studies employ the Q-method. Even though the number of publications each year is still fewer than 10, the use of the Q-method for IS subjects since 2010 regularly appears each year and grows in term of the number of means, compared to years before 2010 (mean of number of publications since 2010 = 2.77 publications; before 2010 = 1.58 publications). The filtering process yielded 20% of papers published in the *Journal of MISQ*, 13.3% in *EJIS*, and 13.3% in *Journal of Management Information Systems*. Other selected journals each have less than 10% (Appendix 1 shows the complete journal distribution). The overview of mapping for 45 publications is listed in Table 1, sorted by the time of publication. Then, following Petersen et al. (2008) to represent the mapping scheme in the form of bubble plots, Figure 2 and Figure 3 show the bubble plot for the systematic mapping based on the percentage of two variables combined.

| Publications The first author (year) | The role of the Q-method | Type of sources for concourse | Comparable DSR process | IS Topic | Study contribution |
|---|---|---|---|---|---|
| Kaplan (1988) | Minor support | 2 without literature | Evaluation | System Analysis & design | Exaptation |
| Kendall (1994) | Major support | 1 not from literature | Solution definition | IS Strategy, management and acquisition | Invention |
| Grover (1995) | Minor support | 1 not from literature | Problem identification | System Analysis & design | Exaptation |
| Tractinsky (1995) | Main method/review | 3 incl. from literature | Solution definition | System Analysis & design | Invention |
| Kettinger (1997) | Minor support | 2 incl. from literature | Solution definition | IS project management | Invention |
| Gottschalk (1997) | Major support | 2 incl. from literature | Problem identification | IS Strategy, management and acquisition | Improvement |
| Moody (1998) | Minor support | 1 not from literature | Evaluation | System Analysis & design | Invention |
| Segars (1998) | Major support | 1 not from literature | Solution definition | IS project management | Improvement |
| Segars (1999) | Minor support | 2 incl. from literature | Solution definition | IS Strategy, management and acquisition | Invention |
| Morgado (1999) | Major support | only from literature | Problem identification | System Analysis & design | Exaptation |
| Gold (2001) | Minor support | only from literature | Problem identification | Data and Information Management | Invention |
| Bhattacherjee (2002) | Minor support | only from literature | Solution definition | System Analysis & design | Invention |
| Thomas (2002) | Main method/review | 2 incl. from literature | Solution definition | System Analysis & design | Exaptation |
| Saeed (2005) | Minor support | only from literature | Evaluation | System Analysis & design | Invention |
| Chang (2005) | Minor support | only from literature | Solution definition | System Analysis & design | Invention |
| Burton-Jones (2006) | Minor support | only from literature | Solution definition | System Analysis & design | Improvement |
| Jahng (2007) | Minor support | only from literature | Evaluation | System Analysis & design | Exaptation |
| Nadkarni (2007) | Minor support | 2 incl. from literature | Evaluation | System Analysis & design | Exaptation |
| Sutton (2008) | Minor support | 1 not from literature | Problem identification | Enterprise architecture | Improvement |
| Klaus (2010) | Main method/review | 1 not from literature | Evaluation | Enterprise architecture | Invention |
| Techatassanasoontorn (2010) | Major support | 1 not from literature | Solution definition | System Analysis & design | Invention |
| Smith (2011) | Minor support | only from literature | Evaluation | System Analysis & design | Exaptation |
| Lu (2011) | Minor support | only from literature | Solution definition | IT Infrastructure | Invention |
| Luo (2012) | Minor support | 2 incl. from literature | Evaluation | System Analysis & design | Invention |
| Sun (2012) | Major support | 2 incl. from literature | Evaluation | System Analysis & design | Invention |
| Wolf (2012) | Minor support | only from literature | Solution definition | IS Strategy, management and acquisition | Exaptation |
| Messerschmidt (2013) | Minor support | only from literature | Problem identification | IT Infrastructure | Exaptation |
| Ou (2014) | Minor support | 2 incl. from literature | Evaluation | System Analysis & design | Improvement |
| Gerlach (2015) | Minor support | 1 not from literature | Evaluation | System Analysis & design | Improvement |
| Grgecic (2015) | Minor support | 1 not from literature | Solution definition | System Analysis & design | Improvement |
| Gerow (2015) | Major support | only from literature | Solution definition | IS Strategy, management and acquisition | Invention |
| Campbell (2015) | Major support | not mentioned | Evaluation | System Analysis & design | Exaptation |
| Benlian (2015) | Major support | 1 not from literature | Evaluation | System Analysis & design | Exaptation |
| Wisniewski (2016) | Minor support | only from literature | Evaluation | System Analysis & design | Exaptation |
| Gautier (2016) | Main method/review | 2 without literature | Evaluation | System Analysis & design | Exaptation |
| Laumer (2016) | Major support | only from literature | Evaluation | System Analysis & design | Improvement |
| Vitari (2016) | Major support | 2 incl. from literature | Evaluation | Data and Information Management | Exaptation |
| Gefen (2017) | Minor support | 2 incl. from literature | Solution definition | System Analysis & design | Exaptation |
| Sarkar (2017) | Main method/review | 3 incl. from literature | Solution definition | IS Strategy, management and acquisition | Invention |
| Laumer (2017) | Major support | 2 incl. from literature | Evaluation | System Analysis & design | Invention |

| Mettler (2017) | Main method/review | 3 without literature | Solution definition | System Analysis & design | Invention |
|---|---|---|---|---|---|
| Wingreen (2018) | Main method/review | 3 incl. from literature | Solution definition | IS Strategy, management and acquisition | Invention |
| Mettler (2018) | Main method/review | 2 incl. from literature | Solution definition | System Analysis & design | Invention |
| Walther (2018) | Minor support | only from literature | Evaluation | System Analysis & design | Improvement |
| Söllner (2018) | Major support | 2 incl. from literature | Evaluation | System Analysis & design | Improvement |

*Table 2.    Overview of the systematic mapping for the Q-method in IS publications.*

The overall mapping results in Table 1 show the significant role of the Q-method in the IS studies as a "support method," with 82.2% (53.3% as minor + 28.9% as major support). This compares to use of the method as the "main method" at only 17.8%. The difference in the role of use is not surprising, since only two journal publications in the list focus on characterizing Q-method use in the field of IS (Mettler et al., 2017; Thomas & Watson, 2002).

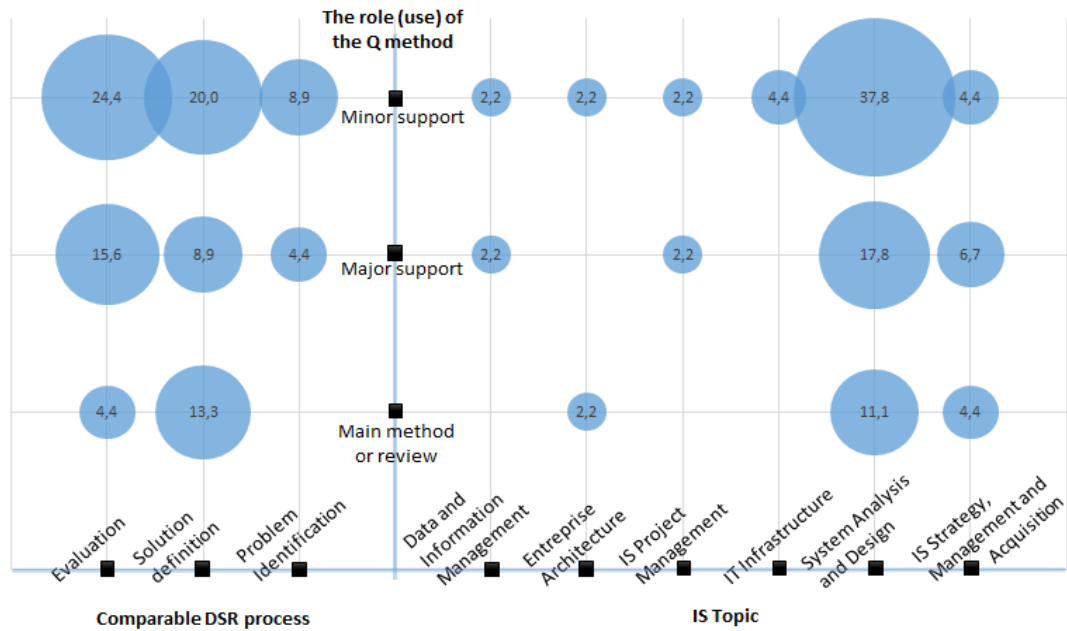

*Figure 2.    Mapping the result by the role of the Q-method, DSR comparable process, and the IS core topics.*

Interestingly, no publications explicitly pointed to the use of DSR in their studies. This is also an indication of the existence of a gap in applying the Q-method in the DSR process. Concerning the type of sources for concourse, the result of overview mapping shows that 68.9% of publications that use the Q-method include scientific literatures or publications as part of the source to collect statements; 48.4% use literature only as the main source; and 51.6% combine the literature with other types of sources. Next, we present the mapping of the result for DSR processes. Figure 2 shows that 24.4% of the papers used the Q-method to support (minor) evaluation studies and 20% aim to define a solution, but no studies use the Q-method as the main method to identify problems. The use of the Q-Method as the main method relates mainly to the process of solution definition.

Based on IS topics in Figure 2, an interesting finding related to publication contribution appears mainly in the IS topic of "system analysis and design," with 37.8% showing the Q-method being used as minor support.

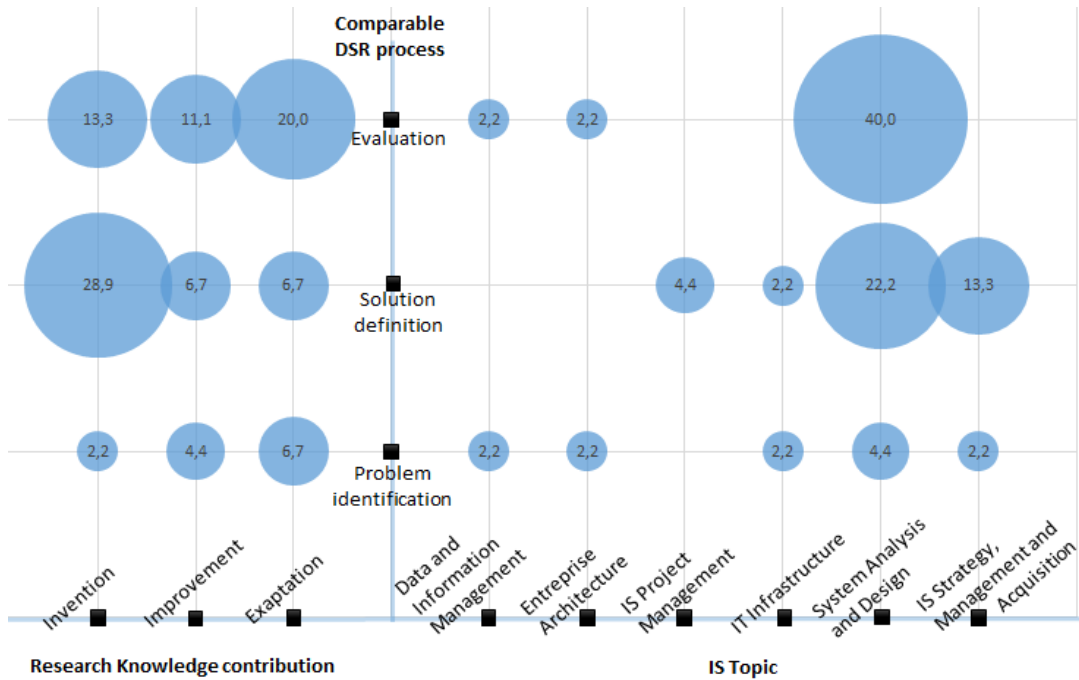

*Figure 3.    Mapping the result by the DSR comparable process, study contributions and the IS core topics.*

Moreover, no significant difference exists between the uses of the Q-method as a major-support or as the "main method" for IS topic "system analysis and design." Furthermore, based on the knowledge contribution of publications, the process of "definition of solutions" that contributes to "the invention" of knowledge is in first place, as shown in Figure 3, with 28.9%, followed by 20% classified as "improvement" for the process "evaluation." Concerning the correlation between the DSR process and the IS topic, 40% of the publications using the Q-method were identified as "evaluation-driven" in the IS topic "the system analysis and design."

## 4    Towards the Integration of the Q-method into DSR Studies

The following analyzes and discusses the systematic-mapping result based on the two guideline questions (GQ1 and GQ2) to answer the main question and explain why the Q-method should be integrated into the DSR process in IS studies.

### 4.1    GQ1: How do IS-researchers use the Q-method?

**The Q-method is commonly used by integrating with another method:** This study shows that the Q-method can be flexibly integrated into IS studies in terms of implementation. IS researchers commonly support the DSR method with other methods—for example, using DSR with Grounded Theory (Gregory, 2011) or with a case study (Nabukenya, 2012). On the one hand, the majority of IS researchers are still not familiar with the Q-method (Wingreen & Blanton, 2018). On the other hand, the mapping shows the use of the Q-method as a support tool for IS researchers was able to provide publications with knowledge contributions in all three essential quadrants of the DSR Knowledge Contribution Framework (Gregor & Hevner, 2013).

**Literature-driven as the basis for the concourse of Q-method for IS studies:** The second interesting result is that the majority of IS studies used the literature review as a starting point for sources in the development of the list of statements, which can also be combined with other sources for the sorting process. Nevertheless, the Q-method has the potential to collect nonscientific digital data, such as social media or blogs (Davis & Michelle, 2011; Lynch et al., 2014; Orchard et al., 2015; Watts & Stenner, 2012). Although IS studies have not fully optimized this potential, found in only two pub-lications (Gefen and Larsen, 2017; Mettler et al., 2017), the future use of the Q-method can also optimize the potential of big data for IS studies (Abbasi, Sarker, & Chiang, 2016). All together (big data as a source of the concourse, the Q-method, and DSR) can become fully implemented as one Q-method for DSR-based IS study.

**Strengthen the result of IS studies related to the system analysis and design:** The third important point related to the Q-method implementation in the IS studies is the high percentage of method use in the analysis and design of the system. That is also the main objective of DSR, i.e., to construct problem-based solutions or reflection-based system improvements (Peffers et al., 2007). The use of the Q-method as the main method also takes second place in the topic of IS strategy design and management. However, the Q-method was also employed as a minor support method for all IS topics. Therefore, this study supports previous studies in using the Q-method as a powerful method for understanding perceptions and representation of societal phenomena (Gautier et al., 2016) in the IS studies. Also, the method can support studies related to technological design (Mettler et al., 2017; Söllner et al., 2018) and different types of IS-topics.

The three answers to GQ1 about the Q-method practice enable an incremental understanding of the Q-method practice in the IS context. The results of the study support earlier studies on the ability of Q-methods to uncover social phenomena in IS (Thomas & Watson, 2002), important for design science. The flexibility of the Q-method both in integrating with other methods and in collecting concourse (also, the less exploited potential of the Q-method to develop concourse from any nonscientific digital data) shows how the potential of the Q-method for integration into different IS topics relates mainly to the analysis of system development. By using the literature as a basis for creating concourse in the Q-method, IS researchers can use the Q-method and deliver knowledge contribution to the IS community by using design sciences based on a study of subjectivity (Cross, 1999). Next is the analysis of the result in response to GQ2.

## 4.2 GQ2: To which DSR processes can the Q-methodology be applied by IS researchers?

**The Q-method can be used to support all the main DSR processes** to answer the question about use of the Q-method for comparable processes in DSR. First, in the identification of problems, challenges or barriers, no publication used the Q-method as the main method. The role of the Q-method is argued only as a supporting tool because the method itself handles the process of concourse development, explained as the main method of a study to identify the problem (Watts & Stenner, 2012). Subsequently, the Q-method was used as a validation (Burton-Jones & Straub, 2006; Jahng et al., 2007; Walther et al., 2018) of the identified problems, in the form of a study proposition with the sorting process in the Q-method. However, on the other hand, the role of the Q-Method in solution determination is a large part of its adoption as the main method, compared to its use in other processes (Figure 2). The importance of determining the best method by user preference (Matzner et al., 2015) increases the appropriateness of using the Q-method as the main method, as this method is reliably achieves consensus, particularly in system design (Figure 3). Therefore, the Q-method can be used as a tool to support research in every critical phase of the DSR process.

**Q-method as the main tool for evaluating IS system based on user preference:** The mapping clearly shows that the Q-method with 66.6% dominates the IS-topic system analysis and design, compared to other IS-topics (see Figure 2). The mapping results support the previous study on the use of Q-methods to explore one's own opinion and reflect one's own experience (Matzner et al., 2015), which plays such an important role in system development. Also, the Q-method has also been used as both main and support method in the evaluation process (Jahng et al., 2007; Nadkarni & Gupta, 2007; Saeed et al., 2005; Smith et al., 2011; Söllner et al., 2018). Therefore, IS researchers may consider the use of the Q-Method as the main tool to support the evaluation process of a system (Figure 3), empirically proven in previous studies as the best method to evaluate a system based on user preferences (Matzner et al., 2015). A practical suggestion is to use the statements in the usability evaluation tools as a source of the concourse, to find consensus on whether a system is usable. By using the Q-method to measure usability, because of the nature of the Q-method, system analysts receive both quantitative and qualitative evaluation results on system usability.

The two answers for GQ2 regarding the comparable DSR process for implementing the Q-method enable an incremental understanding of the Q-method in the DSR process. The Q-Method can be applied to any core DSR process, with an emphasis on the validation and evaluation of propositions and proposed artifacts.

### 4.3   Proposing the integration of the Q-method into a Design-Science-Research study.

This study does not justify the Q-method as the best and only method for the IS community in the DSR process, but rather demonstrates in the mapping result the potential of the Q-method as a complement to the critical process in the DSR. The close link of the Q-method to the human perspective and the existence of genre diversity in DSR (Iivari, 2015; Peffers et al., 2018) speaks to the limitation of the Q-method as well as the strength of the Q method to provide a human-centered IS artifact. The Q-method is not optimal for DSR when a DSR study is based on laboratory experiments or software-based tests, where the influence of the human perspective on the resulting artifact is minimal, such as the simulation-based DSR study in the genre of explanatory design theory (Peffers et al., 2018). Figure 4 summarizes the discussion on the integration of the Q-method, illustrating the proposed framework (including outcomes for each process and integration activities) for the implementation of the Q-method in a DSR study, followed by the explanation.

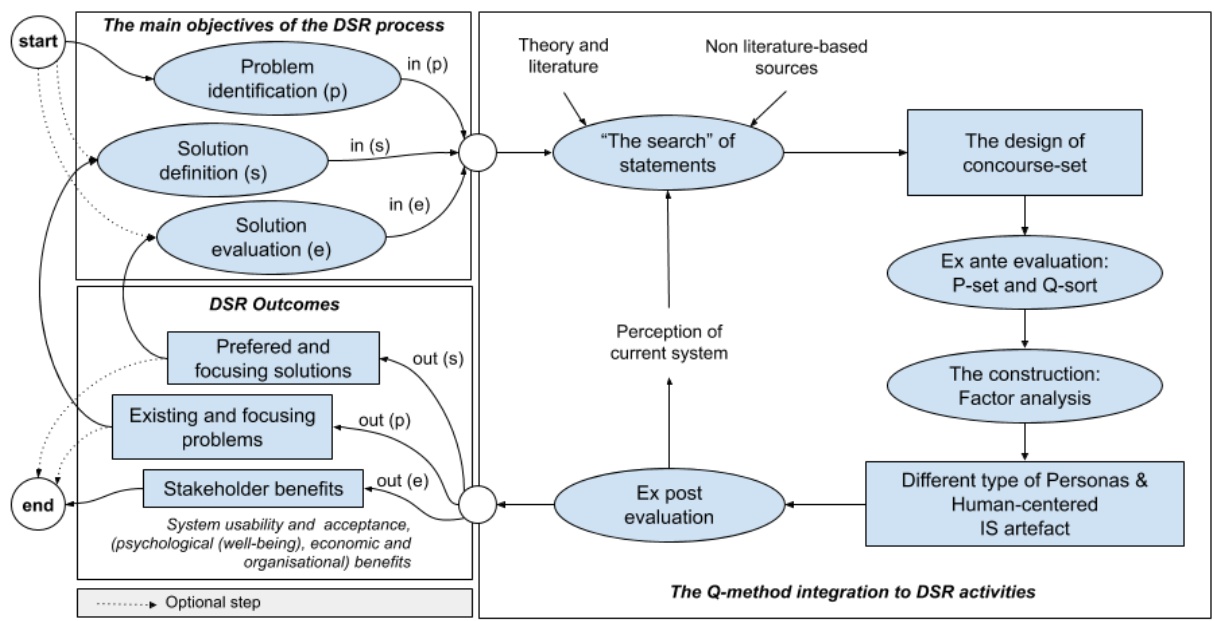

*Figure 4.    Towards the integration of the Q method into a DSR study, modified from (Baskerville et al., 2009).*

Since this study is based on a process view of DSR (Peffers et al., 2007), the Q-method can generally be applied for each genre of DSR (Peffers et al., 2018). However, the integration of the Q method depends on the importance of people being involved in the design of the IS artifact. For the DSR genre of IS design theory, the formation and testing of the proposition becomes the central part (Peffers et al., 2018). The Q-method can help confirm the existence of different statements for a proposition and filter the statements to focus on a particular statement before testing a hypothesis (Burton-Jones & Straub, 2006; Laumer et al., 2017; Saeed et al., 2005). In the DSR Methodology (DSRM), which focuses on the development of applicable artifacts (Peffers et al., 2007; Peffers et al., 2018), the Q method can support all processes, from selecting focus problems and classifying user types based on statements, to selecting the applicable artifact (Mettler & Wulf, 2018; Peffers et al., 2018) by user type and evaluating user preferences (Matzner et al., 2015) that affect both usability and system acceptance (Gefen & Larsen, 2017). In the design-oriented genre of DSR, where the benefit to stakeholders is one of the main topics (Peffers et al., 2018), the Q-method can be used to obtain stakeholders' views on the benefits of the artifact generated by the DSR study. (e.g., the associated benefits of technological interventions for the wellbeing of the user (Mettler et al., 2017; Mettler & Wulf, 2018; Wingreen & Blanton, 2018) which currently can drive the development process of IS design (Pawlowski et al., 2015). While research in DSR with action-design focuses on the reflection and involvement of stakeholders in the creation of an artifact (Peffers et al., 2018; Sein, Henfridsson, Purao, Rossi, & Lindgren, 2011), the Q-Method will be useful both in problem solving and in reflecting the proposed design based on stakeholder opinions.

DSR will affect the design of the concourse of the Q method in which statements in the IS study are formed from two complementary primary sources, namely literature-based, which strongly depend on the use of theory and nonliterary statements. A statement can be in the form of a problem-sentence in the problem-identification process; design principle or propositions in the solution-definition process, and benefits in the form of psychological, economic, or organizational benefits in the evaluation process. For non-literature-based statements including self-reports on experiences with the existing system, digital data collection in the form of large data sets, newspapers, or experts' views can also be used to complement literature-based statements for the concourse. The advancement of the technology and the big-data analysis can serve as a basis for collecting the statements from social media that can be used for the Q-sorting. Besides, in the Q-method, DSR researchers who perform factor analysis can categorize the list of participants (P-set) and identify different gains and pains experienced by the diverse stakeholders. Furthermore, based on the results of systematic mapping and the general process and activities in the Q-method (Watts & Stenner, 2012), which is centralized on the concourse, the integration can be done by focusing on activity level for both DSR and the Q-method. Adopting the episodic DSR represents activities in the DSR process (Baskerville et al., 2009) consisting of search activity, ex ante evaluation, construction activity, and activity related to ex post evaluation.

Moreover, an empirical study proves that the use of the Q-method provides better results compared to rating, ranking, or maximum different scaling assessment method to measure technological preference (Matzner et al., 2015). The use of the Q-method provides a way to fill the gap between quantitative and qualitative debate for the methodological option (Davis & Michelle, 2011) in the design research process. The method can be used as a mixed method with a limited number of respondents (Davis & Michelle, 2011), useful for evaluating the proposed IS artifact in the demonstration process of DSR. Besides, the mapping supports the previous study on the use of the Q-method for theory building (Yang, 2016), as shown in Table 1. The Q-method contributes 44% to the invention of knowledge as one of the main objectives of DSR (Gregor & Hevner, 2013; Gregory, 2011) in the IS community. The Q-Method means not only scores, but also subjectivity, and thus more humanistic investigation (Cross, 2004; Eden, Donaldson, & Walker, 2005) that can discover a balance between freedom and determinism in knowledge interpretation (Dryzek & Braithwaite, 2000). However, the use of the Q-method for DSR depends on the importance of subjectivity for the researchers in their study. The further question to be answered in the DSR-based IS studies is why the Q-method is suitable to include subjectivity.

## 5    Outlook, Limitation and Future Work

This study proposes the use of the Q-method in the DSR process, supplemented by analysis from the results of the mapping on the use of the Q-method in IS studies. It answers the main purpose of the research with a systematic process of mapping that begins with the identification of guideline framework questions. The results of the analysis and discussion of each of the questions are accompanied by an explanation about the study limitations and future research suggestions. Toward this end, IS Scholars are challenged to study and use the Q-method in the DSR-based IS-study to show how the Q-methods can help to generate either rigorous knowledge or applicable artifact for the IS community.

The study design tries to overcome three limitations. First, this study deals only with journals for systematic mapping. To overcome the first limitation, other journal publications were included in the IS field from AIS DL. The publications from AIS DL provided more representation of the Q-method implementation in the IS community. In the future, proceedings papers could be included in a systematic literature review study and use case examples. Secondly, the mapping employed a limited number of variables for the classification. To support the development of a Q-method-based evaluation tool in further study, a different type of IS dimensions could be included (for instance, personal, technological, or organizational dimensions). The third limitation is the selection of the IS core curriculum as a topic for classification. The result of this study provides basic arguments for IS researchers to apply the Q-method for the DSR process in a different area of IS. The IS core topic of system analysis and design still covers a wide range of IS subtopics. Therefore, further study can map the publications based on the subtopic of system analysis and design, and then explore the criteria to implement the Q-method for a different subtopic. The IS subtopic will provide a more detailed picture of the potential corresponding IS topic support by the Q-method.

# 6    Acknowledgment

We would like to thank the IRIS/SCIS community for the conference scholarship, the anonymous reviewers who gave us feedback for improving the paper. The paper is part of a research project funded by the MIWF (Ministry for Innovation, Science and Research of the State of North Rhine-Westphalia) at the positive computing institute of the Hochschule Ruhr West University of Applied Sciences.

# 7    Appendix

Appendix 1: The distribution of the publication in the IS journals that utilized the Q-method.

| Journal | Number of publications | Percentage |
|---|---|---|
| AIS-THCI (AIS Transaction on Human Computer Interaction) | 1 | 2,2 |
| PAJAIS (Pacific Asia Journal of the Association for Information Systems) | 1 | 2,2 |
| SJIS (Scandinavian Journal of Information Systems) | 1 | 2,2 |
| FJMIS (French Journal of Management Information Systems) | 2 | 4,4 |
| ISJ (Information Systems Journal) | 2 | 4,4 |
| JITTA (Journal of Information Technology Theory and Application) | 2 | 4,4 |
| JSIS (Journal of Strategic Information Systems) | 2 | 4,4 |
| CAIS (Communications of the Association for Information Systems) | 3 | 6,7 |
| ISR (Information Systems Research) | 3 | 6,7 |
| JAIS (Journal of the Association for Information Systems) | 3 | 6,7 |
| JIT (Journal of Information Technology) | 4 | 8,9 |
| EJIS (European Journal of Information Systems) | 6 | 13,3 |
| JMIS (Journal of Management Information Systems) | 6 | 13,3 |
| MISQ (Management Information Systems Quarterly) | 9 | 20,0 |

Appendix 2: The distribution of publications in years and quantities

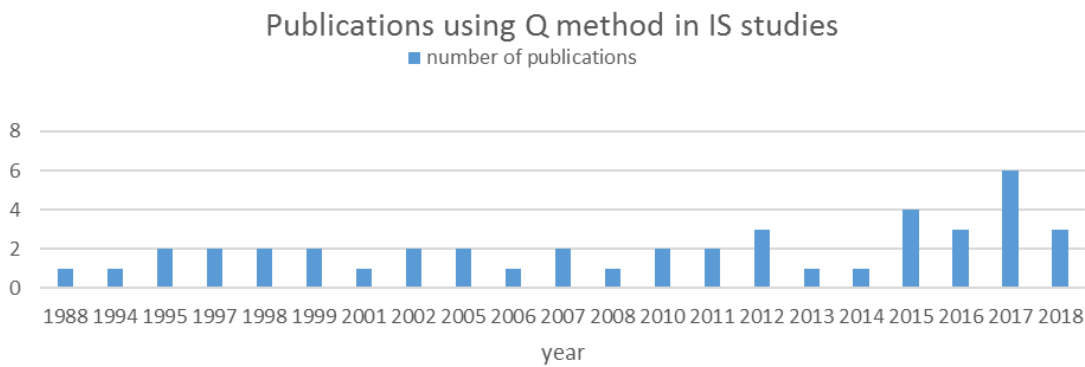